\documentclass[aps,pre,12pt]{revtex4-1}
\pdfoutput=1
\usepackage[utf8]{inputenc}
\usepackage{amsmath}
\usepackage{graphicx}
\usepackage{epstopdf}
\usepackage{comment}
\usepackage{color} 
\usepackage[normalem]{ulem} 
\usepackage{hyperref}

\graphicspath{{Figures/}}

\begin{document}

\newcommand{\eqn}{\begin{eqnarray}}
\newcommand{\eqnend}{\end{eqnarray}}
\newcommand{\kt}{k_B T}
\newcommand{\bs}[1]{\boldsymbol{#1}}
\newcommand{\ps}[1]{\partial_{#1}}
\newcommand{\pare}[1]{\left( #1 \right) }
\newcommand{\corchete}[1]{\left[ #1 \right]}
\newcommand{\fr}[2]{\frac{#1}{#2}}
\newcommand{\wtil}[1]{\widetilde{#1}}
\newcommand{\mc}[1]{\mathcal{#1}}
\newcommand{\ang}[1]{\langle #1 \rangle}
\newcommand{\tex}[1]{\mbox{\scriptsize{#1}}}
\newcommand{\bComment}[1]{{\bf{[#1]}}}
\newcommand{\rComment}[1]{{\bf{\color{red}[#1]}}}
\newcommand{\blueComment}[1]{{\color{blue}#1}}
\newcommand{\what}[1]{\widehat{#1}}

\def\bna{\bs \nabla}
\def\II{\bs I}

\def\dt{\Delta t}
\def\bu{{\bs u}}              
\def\bv{{\bs v}}              
\def\br{{\bs r}}              
\def\bF{{\bs F}}              
\def\dd{\mathrm{d}}

\def\WW{\bs{\mc{W}}}
\def\bk{\bs{k}}
\def\nn{n+1}
\def\n12{n+\fr{1}{2}}
\def\bW{\bs{W}}
\def\bWt{\bs{\wtil{W}}}
\def\bq{\bs{q}}
\def\bw{\bs{w}}

%%%%%%%%%%%%%%%%%%%%%%%%%%%%%%%%%%%%%%%%%%%%%%%%%%%%%%%%%%%%%%%%%%%%%%%%%%%%%%%%
\def\bMtt{\bs{M}} % Donev: Removed _{tt} here to simplify

% Donev macros:
\newcommand{\Donev}[1]{{\bf[{\color{red}#1}]}}
\newcommand{\Delmotte}[1]{{\bf[{\color{blue}#1}]}}
\newcommand{\Balboa}[1]{{{\color{cyan}[#1]}}}
\newcommand{\modified}[1]{#1}
\newcommand{\deleted}[1]{}

\title{Brownian Dynamics of Confined Suspensions of Active Microrollers}

\author{Florencio Balboa Usabiaga}
\affiliation{Courant Institute of Mathematical Sciences, New York
  University, New York, NY 10012, USA}
\author{Blaise Delmotte}
\affiliation{Courant Institute of Mathematical Sciences, New York
  University, New York, NY 10012, USA}
\author{Aleksandar Donev}
\affiliation{Courant Institute of Mathematical Sciences, New York
  University, New York, NY 10012, USA}

\begin{abstract}
\modified{We develop efficient numerical methods for performing many-body Brownian
dynamics simulations of a recently-observed fingering
instability in an active suspension of colloidal rollers sedimented above a wall
[M. Driscoll, B. Delmotte,  M. Youssef, S. Sacanna, A. Donev and P. Chaikin, Nature Physics, 2016, doi:10.1038/nphys3970].
We present a stochastic Adams-Bashforth integrator for the equations of Brownian dynamics, which  has the same cost as but is more accurate than the widely-used Euler-Maruyama scheme, and uses a random finite difference to capture the stochastic drift proportional to the divergence of the configuration-dependent mobility matrix. We generate the Brownian increments using a Krylov method, and show that for particles confined to remain in the vicinity of a no-slip wall by gravity or active flows the number of iterations is independent of the number of particles.
Our numerical experiments with active rollers 
show that the thermal fluctuations set the
characteristic height of the colloids above the wall,
both in the initial condition and the subsequent evolution dominated by active flows.
The characteristic height in turn
controls the timescale and wavelength for the development of the fingering instability.}
\end{abstract}

\maketitle 

%%%%%%%%%%%%%%%%%%%%%%%%%%%%%%%%%%%%%%%%%%%%%%%%%%%%%%%%%%%%%%%%%%%%%%%%%%%%%%%% 
% Introduction 
%%%%%%%%%%%%%%%%%%%%%%%%%%%%%%%%%%%%%%%%%%%%%%%%%%%%%%%%%%%%%%%%%%%%%%%%%%%%%%%% 
\section{Introduction}
Colloidal particles suspended in a liquid often sediment towards a
bottom wall due to gravity, or are confined to move between two parallel walls. 
For example, quasi-two-dimensional experiments use optical microscopy
\cite{Santana-Solano2005, BoomerangDiffusion} or light scattering
techniques \cite{Lan1986} to observe colloidal particles confined
between microscopes slips. 
Synthetic active colloids are usually metallic \cite{Takagi2014, DaviesWykes2016} or
have metallic components \cite{Hematites_Science, Rollers_NaturePhys}, making
them much denser than the surrounding fluid and therefore prone to
sediment towards the floor.
Even neutrally buoyant active colloids are often advected toward the boundaries
by the active flows \cite{Berke2008}.
If the walls are well separated and the
colloids remain much closer to one of the walls, one can model the system as a colloidal suspension in a
half-space bounded by an infinite no-slip wall \cite{StokesianDynamics_Wall}.  This
framework is useful to study both passive \cite{Michailidou2009} and
active suspensions \cite{Berke2008,Galerkin_Wall_Spheres}.

Despite the large number of experiments dealing with colloidal
suspensions close to a single wall, there are few computational
many-body simulations in this geometry.  For active
suspensions, Spagnolie et
al. \cite{Spagnolie2012} and Lushi et al \cite{Lushi2016} both
developed minimal models of swimmer cells (an axisymmetric
swimmer and a bi-flagellated algae respectively) with consistent
hydrodynamic interactions with a nearby wall.  However, they did not
include thermal fluctuations nor considered many-body hydrodynamic interactions.
In a previous work we consider many body simulations of arbitrary
shaped passive or active colloids near a planar wall using a rigid multiblob
method \cite{RigidMultiblobs}, however, we did not discuss how to include Brownian motion.
Recently, Singh and Adhikari
\cite{Galerkin_Wall_Spheres} have used a Galerkin boundary integral representation for
active spheres \cite{BoundaryIntegralGalerkin} to study the crystallization of a two-dimensional
rotating suspension of spheres near a boundary; similar crystal behavior has been
observed in experiments with bacteria and synthetic active colloids \cite{Petroff2015, Hematites_Science},
as well in simulations of unconfined rotating spheres \cite{Lushi2015,Yeo2015}.
These studies neglect all Brownian motion
because their magnitude is estimated to be small compared with the active forces. 
Other works that focus on active suspensions have been
published but in general they do not include Brownian fluctuations or
many-body hydrodynamic interactions \cite{Ishimoto2014, Mathijssen2015, Sambeeta2015}.

For passive suspensions, Michailidou et
al. \cite{Michailidou2009} and Lele et al. \cite{Lele2011} 
combined experiments with Stokesian Dynamics (SD) simulations
\cite{Brady1988} to study colloidal suspensions close to a wall.
Michailidou et al. \cite{Michailidou2009} studied the diffusive
behavior of concentrated colloidal suspension through light scattering
methods; however, they only used SD simulations to solve the mobility
problem in a two particle system and they did not consider Brownian
motion.  Lele et al. \cite{Lele2011} investigated the diffusive modes
of a small colloidal cluster (formed by seven particles) using optical
microscopy.  Their SD simulations included hydrodynamic interactions
\cite{Brady1988,StokesianDynamics_Wall} and Brownian motion \cite{StokesianDynamics_Brownian} but
only for their seven-particle cluster.

It is difficult to perform large-scale
Brownian Dynamics (BD) with hydrodynamic interactions for particles confined near an infinite planar no-slip boundary
using existing numerical methods.
The first key difficulty is efficiently
generating the particles' Brownian displacement. Specifically, it is 
necessary to generate random displacements with a covariance  
proportional to the hydrodynamic \emph{mobility matrix} as to obey fluctuation-dissipation 
balance. Fixman proposed an iterative method to
generate this Brownian noise. His key idea was to realize that it is
enough to obtain an approximation to the noise with a tolerance much
looser than the machine precision. He proposed to use Chebyshev 
polynomials to approximate the Brownian noise \cite{BD_Fixman_sqrtM}.
Fixman's method has become extremely popular in the chemical
engineering community despite its known shortcommings.
Importantly, Fixman's method requires knowing
the largest and smallest eigenvalues of the mobility matrix, which 
requires to use another iterative method \cite{Jendrejack2000}. 
Moreover, Fixman's method often does not 
converge uniformly and the number of iterations to achieve a 
given tolerance generally increases with the number of particles
\cite{SquareRootKrylov}.
\modified{
For periodic or confined domains one can use a fluid solver to compute
the action of the mobility on the fly \cite{Hernandez-Ortiz2009,Chilukuri2014,BrownianBlobs}.
One can combine this
technique with the Fixman's iterative method \cite{BrownianDynamics_OrderN} to generate the Brownian displacements. 
However, this approach requires a Stokes solve every iteration
and it is therefore expensive.
One can instead use fluctuating hydrodynamics \cite{FluctuatingFCM_DC,SpectralRPY,BrownianBlobs}.}
In this approach one first generates a random velocity consistent with the 
fluctuating steady Stokes equations on a grid, and then one interpolates this random velocity to the particle
positions \cite{BrownianBlobs, FluctuatingFCM_DC}.
However, fluctuating hydrodynamics is not straightforward in the geometry of interest here
since it is not possible to use an infinite grid to cover the
half-space above the wall.

To reproduce the equilibrium Gibbs-Boltzmann distribution it is necessary to 
introduce in the BD equations a well-known stochastic drift term proportional to the divergence of the 
mobility matrix \cite{BrownianBlobs,FluctuatingFCM_DC}. Doing this efficiently and accurately is a second key difficulty
in performing Brownian dynamics, especially in non-periodic domains \cite{BrownianBlobs}.
Again, Fixman proposed a method to compute this contribution. His idea was to use a midpoint 
temporal integrator to approximately include the drift term 
\cite{BD_Fixman}. However, the midpoint scheme of Fixman not only involves the
square root of the mobility but also the square root of its inverse, the
resistance matrix. Since there is no pairwise approximations to the resistance matrix even considering 
only far field hydrodynamic interactions \cite{Durlofsky1987}, 
solving resistance problems is generally notably harder than solving mobility problems \cite{RigidMultiblobs,libStokes}.
This makes Fixman's midpoint scheme expensive in practice  \cite{ForceCoupling_Fluctuations},
especially in unbounded systems where the mobility matrix is ill-conditioned.

Here we combine two techniques developed in
recent years to overcome the challenges of performing many-body Brownian simulations
for particles confined above a no-slip boundary. 
We include the effect of the thermal drift using \emph{random finite differences}
(RFD) \cite{BrownianBlobs, BrownianMultiBlobs}.
To generate random displacements with the correct covariance, we use an efficient
iterative method based on Krylov polynomial approximations \cite{SquareRootKrylov, RPY_FMM, SquareRootPreconditioning}.
We find that the screening of hydrodynamics by the nearby boundary makes the 
hydrodynamic interactions sufficiently short-ranged to make the number of
iterations required to achieve a given tolerance \emph{independent} of the number of particles,
making our method scalable to hundreds of thousands of particles. 
In this work we only include hydrodynamic interactions in a far
field approximation and do not use a preconditioner \cite{SquareRootPreconditioning}.
However, it is important to note that both the RFD and the Krylov method are
general techniques that can effectively be used with other methods such as the
boundary integral method of Adhikari et al. \cite{BoundaryIntegralGalerkin, Galerkin_Wall_Spheres}
or our rigid multiblob method \cite{RigidMultiblobs}, with an appropriate preconditioner.

In a recent publication some of us performed simulations with
thousands of colloidal particles to study a fingering hydrodynamic instability
in a suspension of active rollers \cite{Rollers_NaturePhys}.  In a
series of experiments, magnetic colloids roll above a floor with a
specific angular frequency set by an external magnetic field.  The
flow created by the rotation of the colloids creates a strong
collective motion.
After the formation of a shock front of finite width,
a fingering instability appears in the direction transverse to the particles'
motion with the fingers traveling faster than the rest of the suspension.
In the simulations we observed that these fingers can
detach from the suspension and form persistent
motile structures of a well defined size, termed ``critters'',
which move above the wall at
high speed compared with a single particle. Quite remarkably, the
shock, fingers and clusters are formed solely due to hydrodynamic interactions in the presence of the bottom wall.
In our previous work we did not include the effect of Brownian
motion. While the P{\'e}clet number in this active suspension is large, $\text{Pe} \sim 30$,
the effect of thermal noise is key since it sets the characteristic height of the
colloids above the no-slip wall. Moreover, Brownian motion will tend
to break the stable hydrodynamic clusters observed in the absence of
fluctuations.  In this paper we continue the study of this active
suspension by including the effect of thermal fluctuations in
simulations with approximately thirty thousand particles.

\modified{
In Sec. \ref{sec:brownianDynamics} we present the equations of Brownian dynamics
and the tools to solve them efficiently. Then, in
Sec. \ref{sec:rollers} we study the effects of Brownian motion in an active suspension of Brownian rollers. We
conclude the paper with a summary and discussion in
Sec. \ref{sec:conclusions}.
Some technical results are discussed in the Appendices.
}

%%%%%%%%%%%%%%%%%%%%%%%%%%%%%%%%%%%%%%%%%%%%%%%%%%%%%%%%%%%%%%%%%%%%%%%%%%%%%%%% 
% Brownian dynamics 
%%%%%%%%%%%%%%%%%%%%%%%%%%%%%%%%%%%%%%%%%%%%%%%%%%%%%%%%%%%%%%%%%%%%%%%%%%%%%%%% 
\section{Brownian dynamics of Confined Suspensions}
\label{sec:brownianDynamics} 

The Brownian dynamics of $N$ colloidal
particles with coordinates $\bq(t) = \{\bq_1(t),\dots,\bq_N(t)\}$ immersed in a viscous
fluid can be described with the Ito stochastic equation
\eqn
\label{eq:BD} \fr{\dd \bq}{\dd t} = \bMtt \bF + \sqrt{2\kt}
\bMtt^{1/2} \WW + \left(\kt\right) \ps{q} \cdot \bMtt, \eqnend
where $\bMtt(\bq)$ is the
mobility matrix that couples particles (translational) velocities to applied forces
$\bF(\bq;t)$, $k_B$ and $T$ are the Boltzmann constant and the temperature,
and $\WW(t)$ is a vector of independent white noise processes with zero mean and
covariance $\ang{\WW(t) \WW^T(t')} = \II \delta(t-t')$. 
The ``square root'' of the mobility matrix is any matrix (not necessarily square) that
ensures that the Gaussian noise has a covariance proportional to the
mobility matrix, i.e., that the \emph{fluctuation-dissipation balance}
condition
\eqn 
\bMtt^{1/2} \pare{\bMtt^{1/2}}^T = \bMtt,
\eqnend 
holds for all $\bq$.
Equation \eqref{eq:BD} has been used frequently to simulate colloids
\cite{Brady1988} and polymers \cite{Jendrejack2000}, however, it
presents some challenges if one wants to solve it efficiently for
thousands of particles interacting hydrodynamically with a nearby
wall.  

\subsection{Temporal integrators}
\label{subsec:integrators} 

Second order weakly accurate temporal
integrators have been proposed for stochastic equations with
multiplicative noise, see for example Refs. \cite{Cao2002,
WeakSecondOrder_RK}.  However, these schemes are too expensive to
solve \eqref{eq:BD} for tens to hundreds of thousands of particles.
Therefore, we focus here on weakly
first order accurate methods, the simplest of which is the Euler-Maruyama
(EM) scheme, 
\eqn
\label{eq:EM} \bq^{\nn} &=& \bq^n + \dt \bMtt^n \bF^n +
\sqrt{2\kt\dt} \pare{\bMtt^n}^{1/2} \bW^n \\ 
&& + \dt
\fr{\kt}{\delta}\left[\bMtt\pare{\bq^n+\fr{\delta}{2}\bWt^n}\bWt^n
%\right. \nonumber \\&& \left. 
- \bMtt\pare{\bq^n-\fr{\delta}{2}\bWt^n}\bWt^n \right], \nonumber
\eqnend 
where $\dt$ is the time step size, superscripts indicates the point at which
quantities are evaluated (e.g., $\bq^n = \bq(t=n\dt)$ and 
$\bMtt^n=\bMtt\pare{\bq^n}$),
$\delta$ is a small length compared with the particle radius $a$, and
$\bW^{n}$ and $\bWt^{n}$ are uncorrelated vectors of independent
identically distributed (i.i.d.) standard normal variates.

The last term in Eq. \eqref{eq:EM} is a centered \emph{random finite difference} 
(RFD) approximation to the stochastic drift term
that is equal in expectation to $\dt \kt \left(\ps{q} \cdot \bMtt\right)^n$
for sufficiently small $\delta$; one can also use a one-sided difference.
The RFD term guarantees that the EM scheme is a consistent
integrator of \eqref{eq:BD} \cite{BrownianBlobs,BrownianMultiBlobs},
but is simpler and more efficient in practice
than the Fixman midpoint scheme.
Assuming that the product between the mobility matrix $\bMtt$ and a force vector
is computed with accuracy $\varepsilon$,
and denoting the radius of the particles with $a$,
a balance between truncation and
roundoff error in the centered RFD is achieved for $\delta/a\sim\varepsilon^{1/3}$
($\delta/a\sim\varepsilon^{1/2}$ for one-sided difference);
choosing $\delta/a=10^{-6}-10^{-5}$ is appropriate for the simulations reported here.

\begin{comment}
A slightly cheaper alternative is to use a one-sided RFD of the form
\eqn
\dt \fr{\kt}{\delta}\corchete{\bMtt\pare{\bq^n+\delta\bWt^n}\bWt^n
- \bMtt\pare{\bq^n}\bWt^n},
\eqnend 
since computing the matrix-vector product $\bMtt^n\bWt^n$ can be combined with
computing the matrix-vector product $\bMtt^n \bF^n$. Since in our case the total number of matrix-vector products
per time step will turn out to be on the order of a dozen (see Section \ref{subsec:BrownianNoise}),
the cost saving of the one-sided difference is negligible and we use the
centered difference in the simulations reported here. 
\end{comment}

The EM scheme \eqref{eq:EM} is not particularly accurate even for
deterministic equations. Another first
order weakly accurate scheme that we empirically find to give notably
better accuracy is the stochastic Adams-Bashforth (AB) scheme, 
\eqn
\label{eq:AB} \bq^{n+1} &=& \bq^n + \dt \corchete{\fr{3}{2}\bMtt^n
\bF^n - \fr{1}{2} \bMtt^{n-1}\bF^{n-1}} 
+ \sqrt{2\kt\dt} \pare{\bMtt^n}^{1/2} \bW^n \\  
&& + \dt
\fr{\kt}{\delta}\left[\bMtt\pare{\bq^n+\fr{\delta}{2}\bWt^n}\bWt^n
% \right. \nonumber \\ && \left. 
- \bMtt\pare{\bq^n-\fr{\delta}{2}\bWt^n}\bWt^n\right]
\nonumber .
\eqnend 
Here we use the same Brownian displacement as in \eqref{eq:EM} but we evaluate the deterministic
forces with a second order method.
\modified{Observe that the AB and EM schemes have the same
computational cost per time step, with the only difference being that the AB scheme requires storing one more vector.
We are not aware of an analysis of the weak accuracy of multistep schemes such as \eqref{eq:AB},
however, we demonstrate empirically in Appendix \ref{sec:accuracy} that
the AB scheme improves not only the deterministic but also the stochastic accuracy for time step sizes
that resolve all relevant physical time scales.}
One could build other integrators
based on the mid-point or the trapezoidal rules \cite{DFDB}
at the cost of increasing the cost per time step.
\modified{We have found that the AB scheme has a similar tradeoff of accuracy
versus computational cost as two steps Runge-Kutta methods. We
we use the AB scheme here because it is more efficient in the deterministic setting.}

\subsection{Mobility matrix}
\label{subsec:mobility} 
To make \eqref{eq:BD} well-posed, we need a
mobility matrix $\bMtt$ that is symmetric positive
semidefinite (SPD) for all particle configurations. In the presence of an infinite
no-slip wall the mobility should include the effect of the wall on the
hydrodynamic interactions.  We use a standard pairwise approximation that
includes the long-range contribution. For spherical particles of
radius $a$ we can write the pair mobility matrix between two different
particles $i$ and $j$ using the generalized Rotne-Prager (RP)
tensor $\bs{\mathcal{R}}$ \cite{StokesianDynamics_Wall,RPY_Shear_Wall},
\eqn
\label{eq:mobilityDifferential} 
{\bMtt}_{ij} = \bs{\mathcal{R}} \left( \bq_i, \bq_j \right) =
\pare{\II + \fr{a^2}{6}\bna^2_{x}}\pare{\II + \fr{a^2}{6}\bna^2_{y}}
\bs{G}(\bs{x}, \bs{y}) \bigg|_{\bs{x}=\bq_i}^{\bs{y}=\bq_j} , 
\eqnend 
where $\bs{G}(\bs{x}, \bs{y})$ is the Green's function of the Stokes 
equation with the appropriate boundary conditions, and the parentheses 
are the Fax{\'e}n operators. Blake found the Green's function for the 
Stokes problem in a half space using a method of images 
\cite{Blake1974}, and Swan and Brady have used this to compute explicit expressions for the 
RP tensor \eqref{eq:mobilityDifferential} in the presence of a wall \cite{StokesianDynamics_Wall}.

A practical problem of the mobility given by Swan and Brady
\cite{StokesianDynamics_Wall} is that it only remains positive semidefinite if the
particles do not overlap with each other or with the wall.  The
no-overlap requirement seems natural; however, to completely avoid
overlaps in a BD simulation it is necessary to use very strong
repulsive potentials and therefore small time step sizes. We prefer to use
softer potentials that allow for larger time step sizes at the cost of an
occasional overlap; therefore, we need to generalize the mobility
matrix to have a positive semidefinite expression for all particle
configurations.

For particles overlapping with each other, instead of applying the
Fax{\'e}n operators to the Green's function as in Eq. \eqref{eq:mobilityDifferential}
one can compute the mobility in two steps. The first step is to compute the flow $\bv(\br)$
created by a force density $\bF / (4\pi a^2)$ acting on the surface $S_i$ of
one of the spheres. The second step is to compute the velocity of the second particle as
the average of the velocity flow $\bv(\br)$ over the surface $S_j$ of the other particle.
This leads to the integral form of the Rotne-Prager-Yamakawa (RPY) tensor \cite{RPY_Shear_Wall},
\eqn
\label{eq:mobilityOverlap} 
\bs{\mathcal{R}} \left( \bq_i, \bq_j \right) = \fr{1}{(4\pi a^2)^2}
\oint_{\bs{x}\in S_i} \oint_{\bs{y}\in S_j} \bs{G}(\bs{x},\bs{y}) \, dS_i(\bs{x}) dS_j(\bs{y}),
\eqnend 
which is identical to the differential form \eqref{eq:mobilityDifferential} for non-overlapping particles.
For particles that overlap each other but do not overlap the wall,
it has been shown that the correction relative to \eqref{eq:mobilityDifferential}
is independent of the boundary conditions \cite{RPY_Shear_Wall},
and therefore it is the same as the one obtained by Rotne and Prager in an
unbounded domain \cite{RotnePrager}.
Specifically, the RPY mobility \eqref{eq:mobilityOverlap} in the presence of a wall
that we use in this work is given by adding the wall-corrections
given in Eqs. (B1) and (C2) in \cite{StokesianDynamics_Wall} to the RPY tensor in an unbounded domain
given in Eq. (3.12) in \cite{RPY_Shear_Wall}.
Note that the self-mobility of a
particle is given by \eqref{eq:mobilityOverlap} with $j=i$.
Equation \eqref{eq:mobilityOverlap} cannot be applied directly when one or both
of the particles overlap the wall because part
of the particles surface extends beyond the wall. 
\modified{In Appendix \ref{sec:SPDmobility} we explain how we
regularize the mobility matrix for particles overlapping the wall
so that $\bMtt$ is SPD for all particle configurations.}

We use a simple direct summation implemented in PyCUDA \cite{PyCUDA},
which runs in Graphical Processing Units (GPUs), to compute the matrix
vector product $\bMtt \bF$.  Although the computational cost of this
method is formally $\mc{O}(N^2)$ in the number of particles $N$, the GPUs allow for a high
parallelization of the calculation making this part of the algorithm
fast enough to be used with thousands of particles \cite{DirectStokesian_GPU}.
We want to point out that in certain circumstances
the matrix vector product can be computed in a quasi-linear time in
the number of particles. For example, for periodic domains one can
compute the matrix vector product using Ewald splitting methods
\cite{SpectralRPY}, and for unbounded domains one can use Fast Multipole Methods (FMM) \cite{RPY_FMM}.
For the half-space problem it is also known how to use FMM in the limit
$a \to 0$ (i.e. for Stokeslets) \cite{OseenBlake_FMM}, but not for $a>0$, to our knowledge.
When deciding which method should be used one has to bear in mind that
often, a simple GPU implementation can be faster than a method with a
better theoretical scaling unless the number of particles is above about a
hundred thousand for present-day hardware and FMM implementations \cite{RigidMultiblobs}.
\deleted{Our PyCUDA codes are publicly available at
\url{https://github.com/stochasticHydroTools/RigidMultiblobsWall}.}

\subsection{Brownian noise}
\label{subsec:BrownianNoise} 

\begin{figure}
\includegraphics[width=0.99 \columnwidth]{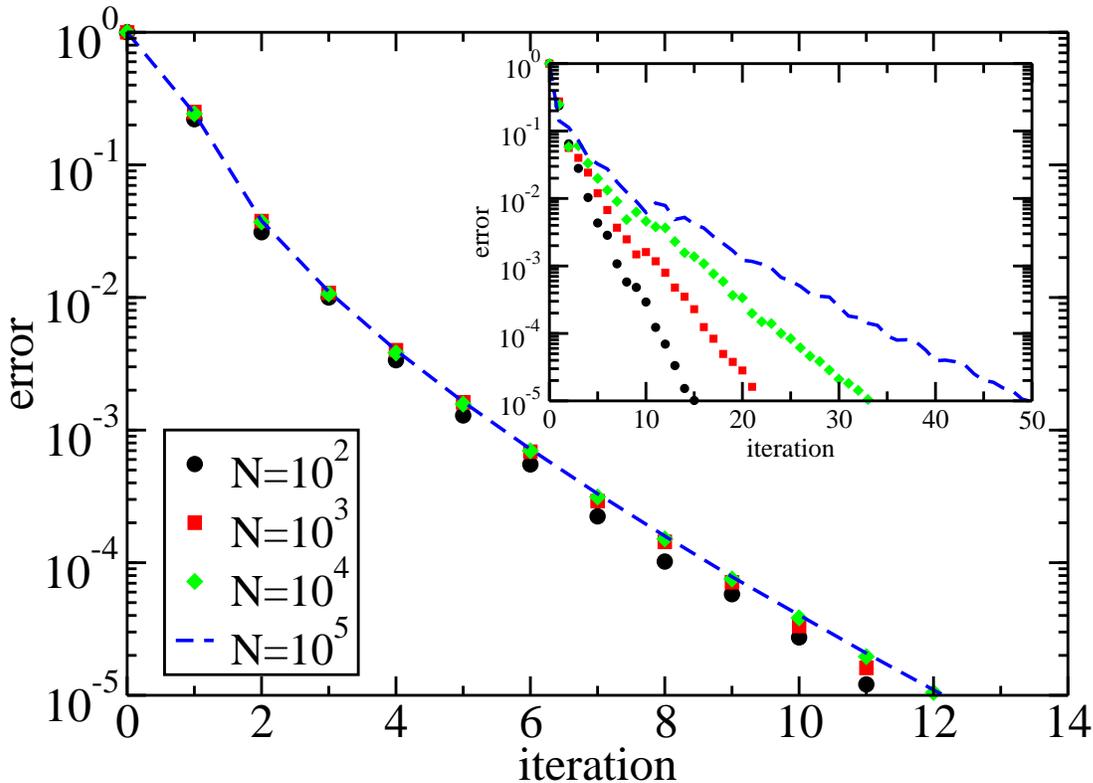}
\caption{\modified{Error versus iteration during the generation of the
noise with the Lanczos algorithm for an equilibrium suspension of particles
sedimented over the wall, for different number of particles
$N$, keeping fixed the area fraction $\phi=0.25$ and gravitational height
(characteristic height over the wall) $h_g \approx 1.6a$.
\modified{Similar results are obtained for a broad range of physically relevant gravitational heights.
The inset shows the convergence of the Lanczos
method for the same configurations when the screening of the hydrodynamic
interactions by the wall is not included, i.e., the particle-particle
hydrodynamic interactions decay as inverse of the distance.}}
}
\label{fig:residualNoise}
\end{figure}

A key step in integrating the BD equation \eqref{eq:BD}
using the scheme \eqref{eq:EM} or \eqref{eq:AB} is generating the noise term
$\sqrt{2\kt \dt} \bMtt^{1/2} \bW$.
We compute the random term $\bMtt^{1/2}\bW$ with the Krylov method
proposed by Ando et al. \cite{SquareRootKrylov} (see also Section 3 in
Ref. \cite{RPY_FMM}).
In short, we use the Lanczos
algorithm with $m$ iterations to compute the Krylov subspace
$K_{m}(\bMtt,\bW)=\mathrm{span}\{\bW,\,\bMtt\bW,\,\ldots,\,\pare{\bMtt}^{m-1}\bW
\}$ and to approximate the random noise, $\bs{g}_m \approx
\bMtt^{1/2}\bW$, in that subspace.  As in Ando el
al. \cite{SquareRootKrylov} we estimate the relative error $\epsilon_m$ at the $m$th
iteration by the relative difference with the noise generated in the
previous iteration, $\epsilon_m = \|\bs{g}_m - \bs{g}_{m-1}\|_{2} /
\|\bs{g}_{m-1}\|_2$.
\modified{The computational cost of this method is $\mc{O}(m^4 + m \text{MV})$ where
MV denotes the cost of computing one matrix vector product.}

\modified{
Figure \ref{fig:residualNoise} shows the Lanczos algorithm's
convergence for suspensions with different number of particles
sedimented over the wall.
In general, iterative methods do not scale
well for dense suspensions in three dimensions due to the slow $1/r$
decay of the Oseen/RPY tensor.  The condition number of the mobility
matrix increases with the system size which in turn increases the
number of iterations to generate the noise.  For particles in a half
space, however, the situation is much better if the particles are
confined in a thin layer next the wall.  The wall screens the
hydrodynamic interactions between particles, which decay like $1/r^3$
when $r \gg h$ \cite{HI_Confined_Decay}, where $h$ is the particles' height. We find that if
particles are closer to the wall than they are to remote particles,
the condition number is essentially independent of the number of
particles, and in practice the Krylov method converges in a constant
number of iterations.  The results confirm that the convergence of the
Krylov method is essentially independent of the number of particles
and that less than a dozen iterations are required to reach a relative
error tolerance $\sim10^{-5}$
\footnote{\modified{
The fact the mobility matrix is well-conditioned
makes it also possible to use Fixman's midpoint scheme in practice instead of the RFD,
since one can use the Lanczos method to efficiently compute $\bMtt^{1/2} \bW$ and $\bMtt^{-1/2} \bW$  together.}}.
We note that as the particles start to
overlap with the wall the condition number of the mobility matrix
increases, however, the number of iteration is not affected for
typical configurations (not shown).
}

\modified{
We have tried to improve convergence of the Lanczos method
by preconditioning \cite{SquareRootPreconditioning}. A simple block diagonal preconditioner, which
neglects the hydrodynamic interactions between distinct particles
but does include the hydrodynamic interactions of each particle with the wall \cite{RigidMultiblobs},
gives only a very small improvement in the number of iterations.
A preconditioner based on a incomplete Cholesky factorization of a truncated mobility matrix
helps convergence greatly but also significantly increases the memory requirements and the complexity of the code,
and also increases the cost per iteration.
Therefore we do not use preconditioning \cite{SquareRootPreconditioning}
in this work.
}

\modified{
To evaluate the screening effect
of the wall on the Lanczos algorithm we compute the Brownian noise for
the same particles configurations but neglecting the hydrodynamic
interactions with the wall. The inset in Fig. \ref{fig:residualNoise}
shows that the convergence worsens with increasing number of particles. 
}
\modified{
For bulk suspensions either a preconditioned Krylov method \cite{SquareRootPreconditioning}
or a direct (non-iterative) method \cite{SpectralRPY} is needed
to handle the ill-conditioning due to the long-ranged hydrodynamic interactions.
}

\section{Active Suspensions: Brownian Rollers}
\label{sec:rollers}

In this section, we perform Brownian Dynamics simulations of the
fingering instability which was discovered in
\cite{Rollers_NaturePhys}. In that work, it was shown that suspensions
of rollers (spherical particles) rotating parallel to a floor form
narrow shock-like fronts which destabilize in the direction transverse
to their motion. This transverse instability results in the formation
of fingers with dense fingertips that propagate very fast compared to
a single roller.  Deterministic simulations showed that the wavelength
of the fastest growing mode is set by the particles' height above the
wall, which experiments suggest corresponds to the gravitational
height. Here we assess how thermal noise affects the characteristic
wavelength of the instability.

The experimental system consists of spherical colloidal particles of
radius $a=0.656 \, \mu$m in which a small cube of hematite is embedded
\cite{Sacanna2012}. The hematite is a canted anti-ferromagnet and thus
provide the rollers with a small permanent magnetic moment
$|\mathbf{m}| = 5\cdot 10^{-16}$ A$\cdot$m$^2$. The particles are
suspended in water ($\eta = 1$ mPa$\cdot$s) at an average height from
the wall given by the gravitational height $h_g = a + k_BT/mg$, which
is set by the competition between gravity and Brownian motion.
\modified{Here $g$ denotes gravity and $m$ the excess mass of the particles over the
  expelled fluid.}
The particles are rotated with an oscillating magnetic field $\mathbf{B} =
B \left[\cos(\omega t) \hat{\mathbf{x}} + \sin(\omega t)
\hat{\mathbf{z}}\right]$ with frequency $f=\omega/2\pi$ about the
$\hat{\mathbf{y}}$ axis. If the field is strong enough and the
frequency low enough to overcome the viscous torque exerted by the
surrounding fluid, the particles rotate synchronously with
$\mathbf{B}$ at a rate $\omega$.  \deleted{ Otherwise they slip
relative to $\mathbf{B}$ with a varying phase lag. Our previous study
shows that the critical driving frequency is
  \begin{equation} \omega_c = mB/8\pi\eta a^3f(a/h)
  \end{equation} where $f(a/h) =
\left(1-\tfrac{15}{48}\tfrac{a^3}{h^3}\right)^{-1}$ is the leading
order wall correction for the torque-rotation mobility coupling
\cite{StokesianDynamics_Wall}.  The typical value measured in the
experiments for a field strength $B=2.94$ mT is $\omega_c = 170$
rad$/s$.  } In our previous deterministic simulations of the fingering
instability \cite{Rollers_NaturePhys}, we kept the particle angular
velocity fixed at $\omega$, and computed the torque required to
maintain that specified rotation.  However, qualitatively identical
behavior is observed if the particles are driven by a constant torque
instead, which can also be realized experimentally and is somewhat
cheaper to simulate.  Therefore, here we exert a constant identical
torque on every particle $\boldsymbol{T} = 8\pi\eta a^3 \omega
\hat{\mathbf{y}}$, where the driving frequency is in the experimental
range, $\omega = 10 \text{Hz} = 62.8 \text{rad}/s$.

Due to the rotation a single particle will translate (roll) above the
wall in the direction perpendicular to the rotation. Additionally, the
rotation creates a strong flow that can advect nearby particles.  We
can model these active flow effects by including in the right hand
side of \eqref{eq:BD} the deterministic term $\bMtt_C \bs{T}$ where
$\bMtt_C$ is the RPY translation-rotation mobility in the presence of
a wall.  For non-overlapping particles the equivalent of
\eqref{eq:mobilityDifferential} gives the differential form of the RP
translation-rotation mobility, \eqn
\label{eq:mobilityDifferentialTR} \left({\bMtt}_C\right)_{ij} =
\bs{\mathcal{R_C}} \left( \bq_i, \bq_j \right) =
\pare{\II + \fr{a^2}{6}\bna^2_{x}}\pare{\frac{1}{2}\bna_{y}\times}
\bs{G}(\bs{x}, \bs{y}) \bigg|_{\bs{x}=\bq_i}^{\bs{y}=\bq_j}, \eqnend
and for overlapping particles one can use a suitable generalization of
\eqref{eq:mobilityOverlap}, see Eq. (3.16) in \cite{RPY_Shear_Wall}.
The final form of $\bs{\mathcal{R_C}}$ can be computed explicitly by
adding the wall-corrections given in Eqs. (B2) and (C3) in
\cite{StokesianDynamics_Wall} to the corresponding expression for an
unbounded domain given in Eq. (3.16) in \cite{RPY_Shear_Wall}.  In the
temporal integrators we treat the forcing term $\bMtt_C \bs{T}$ like
the deterministic force $\bMtt \bF$.

\modified{
We include a soft pairwise steric interaction $U(r)$ between the particles and between the particles and
the wall to avoid severely limiting the time step size,}
\eqn 
U(r) = \left\{ \begin{array}{cc} 
% \infty & \mbox{if } r < 0, \\
U_0 + U_0 \fr{d-r}{b} & \mbox{if }  r < d, \\ 
U_0 \exp\pare{-\fr{r-d}{b}} & \mbox{if } r \ge d.
\end{array} \right.  
\label{eq:soft_potential}
\eqnend 
For particle-particle interactions, $d=2a$ and $r$ is the distance between the
centers of the particles, while for particle-wall interactions, $d=a$
and $r$ is the distance of the center of the particle to the
wall. The energy scale $U_0$ and interaction range $b$ are parameters that
control the strength and decay (range) of the potential. Even though the 
potential $U(r)$ does not strictly prevent a particle from crossing the wall, we can
make the probability of such events arbitrarily small by setting the
energy of a particle about to cross the wall  
to be very large compared with the thermal energy; in this paper we use $(1 + a/b) U_0
= 44 \kt$ and we never observe a particle escaping the
physical domain in practice.
The potential $U(r)$ introduces a characteristic steric time scale 
\eqn
\label{eq:stericTime} 
\tau_U &=& \fr{6\pi\eta a^2 b}{U_0}.  
\eqnend 
We choose the values $U_0=4\kt$ and $b=0.1a$ to have a small number of overlaps
but keeping $\tau_U$ no more than about an order of magnitude smaller
than the other characteristic times in the system, the diffusive time $\tau_D=3\pi\eta
a^3/(\kt)$ and the sedimentation time $\tau_g=6\pi\eta a^2/(mg)$.

Each simulation contains $N = 2^{15} = 32,768$ particles which are
initialized by sampling the equilibrium Gibbs-Boltzmann distribution
of a two-periodic suspension with a periodic cell in the $x-y$
directions of extent $60a \times 4915a$ using a Monte Carlo method.
The only parameter which is varied between the simulations is the
gravitational height, $h_g = (1.5,\, 3.5,\, 6.1)a$, by changing the
excess mass of the rollers, $m = (1, \, 0.2, \, 0.1) \times 1.27 \cdot
10^{-15}$ kg.  
The total simulation time is $40-160$ s depending on
$h_g$, ensuring that the instability is well-developed for all
gravitational heights. The tolerance of the Lanczos algorithm is $\epsilon_m = 10^{-3}$.
\modified{The time step size is set to $\Delta t = 0.5 \tau_U = 0.016$s 
to ensure that the correct Gibbs-Boltzmann equilibrium distribution is obtained 
for passive suspensions, see Appendix \ref{sec:accuracy}, and}
to ensure that the active motion does not
introduce significant overlaps; we have confirmed that essentially
identical results are obtained by using $\Delta t = 0.008$s (not shown
here).  All the results below were averaged over four realizations for
each case.

\begin{table}[]
  \begin{center}
    \begin{tabular}{ |c|c|c|c| } \cline{2-4} \multicolumn{1}{c|}{} &
$h = 1.5a$ & $h = 3.5a$ & $h = 6.1a$ \\ \hline Deterministic & 9.4 s &
17.3 s & 32.6 s\\ \hline Stochastic & 11.8 s & 26.9 s & 42.9 s \\
\hline
    \end{tabular}
  \end{center}
  \caption{Characteristic time $t^{\star}$ as defined in
\eqref{eq:charac_time}.}
  \label{tab:charac_time}
\end{table}

From the particle positions at a given time $t$ we can compute the
\modified{distribution $P(h)$  of particle heights $h$ above the wall and the}
empirical number density $n(x,y;t)$ projected onto the $x-y$
plane. From $n(x,y;t)$ we compute two quantities of interest. The first
quantity is the distribution of particle positions along the direction
perpendicular to the unstable front, $\rho(x;t)=\int_y
n(x,y;t)dy$. The second quantity is the Fourier transform of
$n(x,y,t)$ at $k_x = 0$, i.e., the Fourier transform
$\hat{n}(k_y=2\pi/\lambda,t)$ of the number density along the
direction of the front.  \modified{In this second case we only use
particles in the front, specifically,
we only include the $70\%$ of the particles with the largest $x$-coordinates.
This ensures that the Fourier modes are not affected by the
particles left behind the shock front. We have confirmed that
essentially the same results are obtained when include between $50\%$ and $90\%$ of the particles.}
In order to compare the stochastic and deterministic simulations for various $h$,
we define a characteristic time $t^{\star}$ as the time where
$\hat{n}(k_y,t)$ reaches its maximum power,
\begin{equation} t^{\star} = \underset{t}{\operatorname{argmax}}
\int_{k_{\min}}^{k_{\max}} |\hat{n}(k_y,t)|^2 dk_y,
  \label{eq:charac_time}
\end{equation} where we choose $k_{\min}=2.0\cdot 10^{-3}$
$\mu$m$^{-1}$ and $k_{\max}=0.26$ $\mu$m$^{-1}$.  Table
\ref{tab:charac_time} compares $t^{\star}$ between the deterministic
and stochastic case.  As shown by the value of $t^{\star}$, the
dynamics in the Brownian system is slower than the deterministic one
by about a factor of $1.4$ for all three values of $h_g$.  This
difference in the time evolution is expected since the Brownian
rollers spend more time farther away from the floor and thus translate
slower.

\begin{figure*}
  \includegraphics[width=0.49
\textwidth]{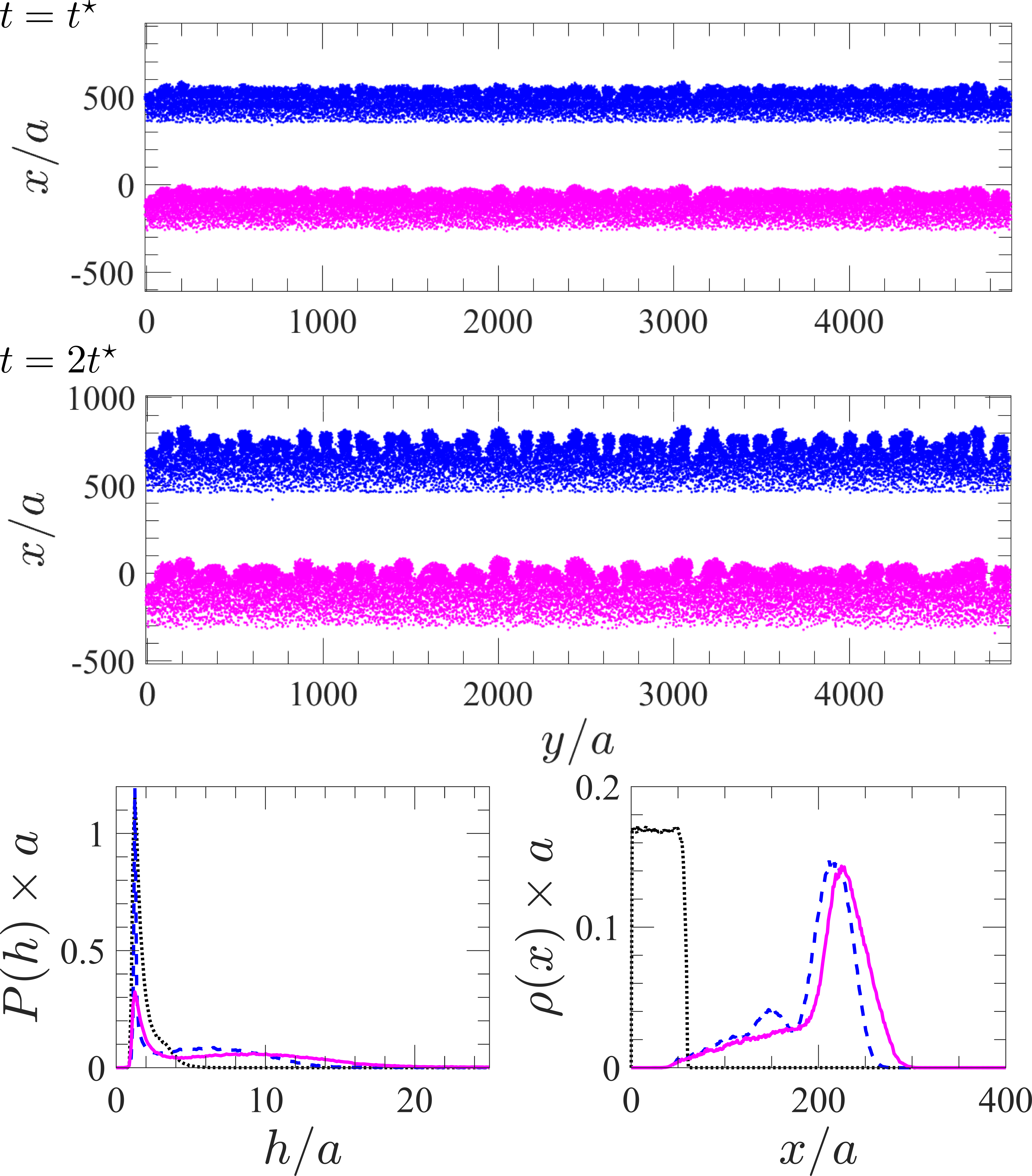}
  \includegraphics[width=0.49
\textwidth]{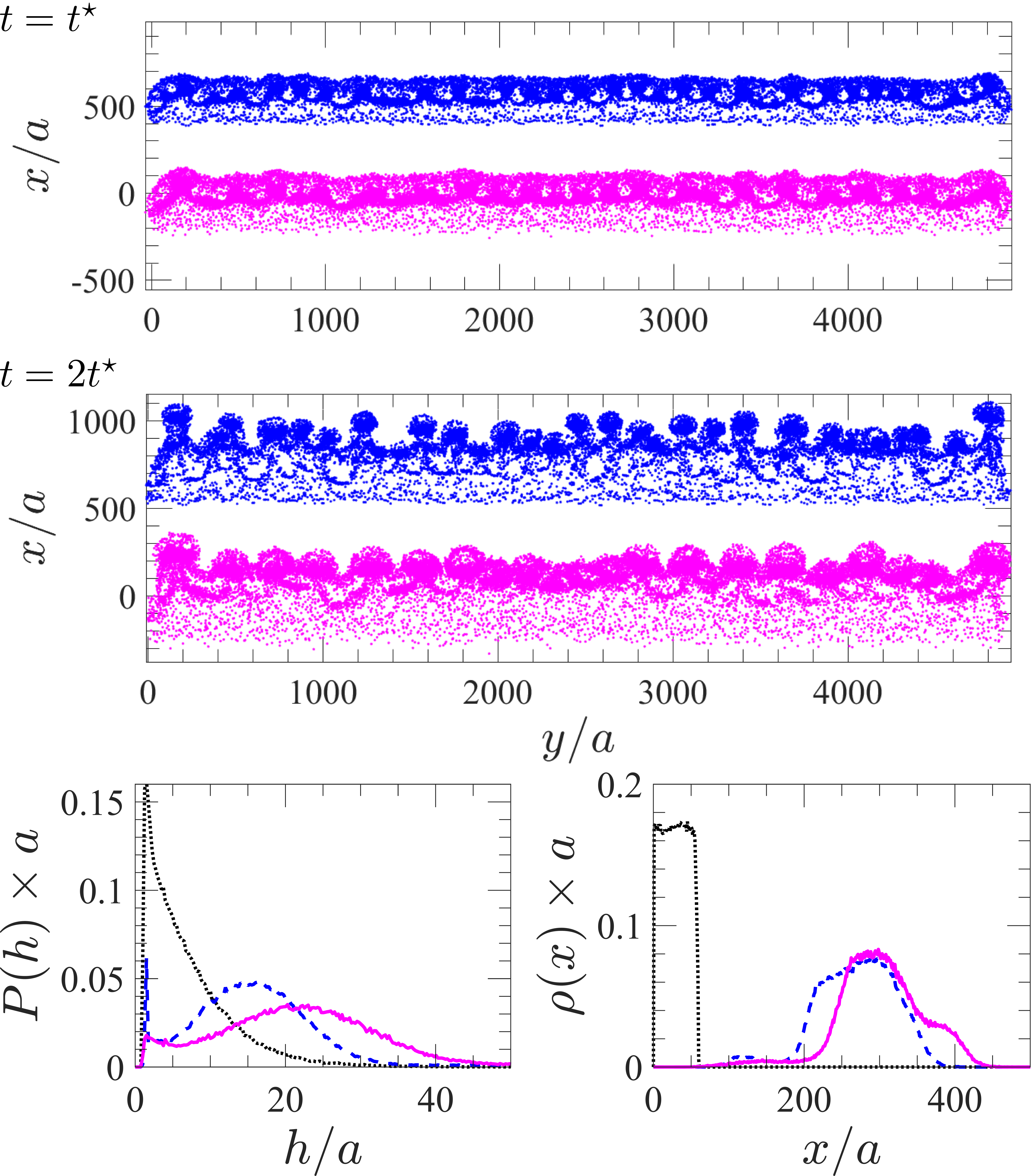}
  \caption{Comparison between deterministic (blue) and stochastic
(magenta) simulations at multiples of the characteristic time
$t=t^{\star}$ given in Table \ref{tab:charac_time} for two
gravitational heights, $h_g = 1.5a$ (left half) and $h_g = 6.1a$
(right half). Top and middle subpanels: particle positions projected
onto the $x-y$ plane (as would be observed in experiments) for a
single realization, at $t=t^{\star}$ (top), when the instability
begins to develop, and $t = 2t^{\star}$ (middle), when fingers have
formed. The deterministic and stochastic results are staggered in the
vertical direction for clarity. Bottom left subpanel: height
distribution $P(h)$ at $t=t^{\star}$ averaged over four realizations.
Bottom right subpanel: particle distribution $\rho(x)$ at
$t=t^{\star}$ averaged over four realizations. In all bottom panels,
the black dotted line show the initial distributions at $t=0$ s.}
  \label{fig:comparison_histograms}
\end{figure*}

\begin{figure*}
  \includegraphics[width=0.50 \textwidth]{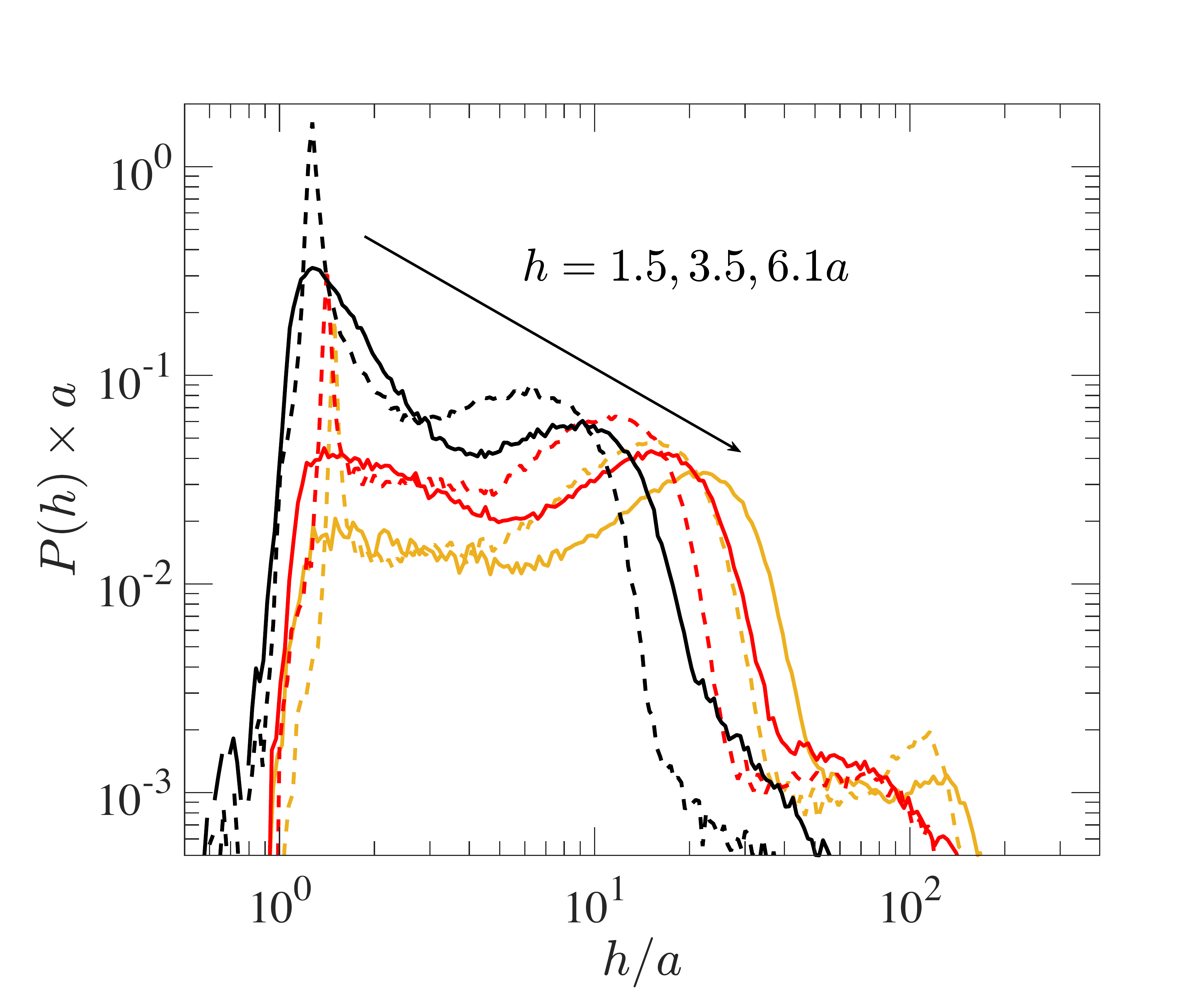}
  \includegraphics[width=0.49 \textwidth]{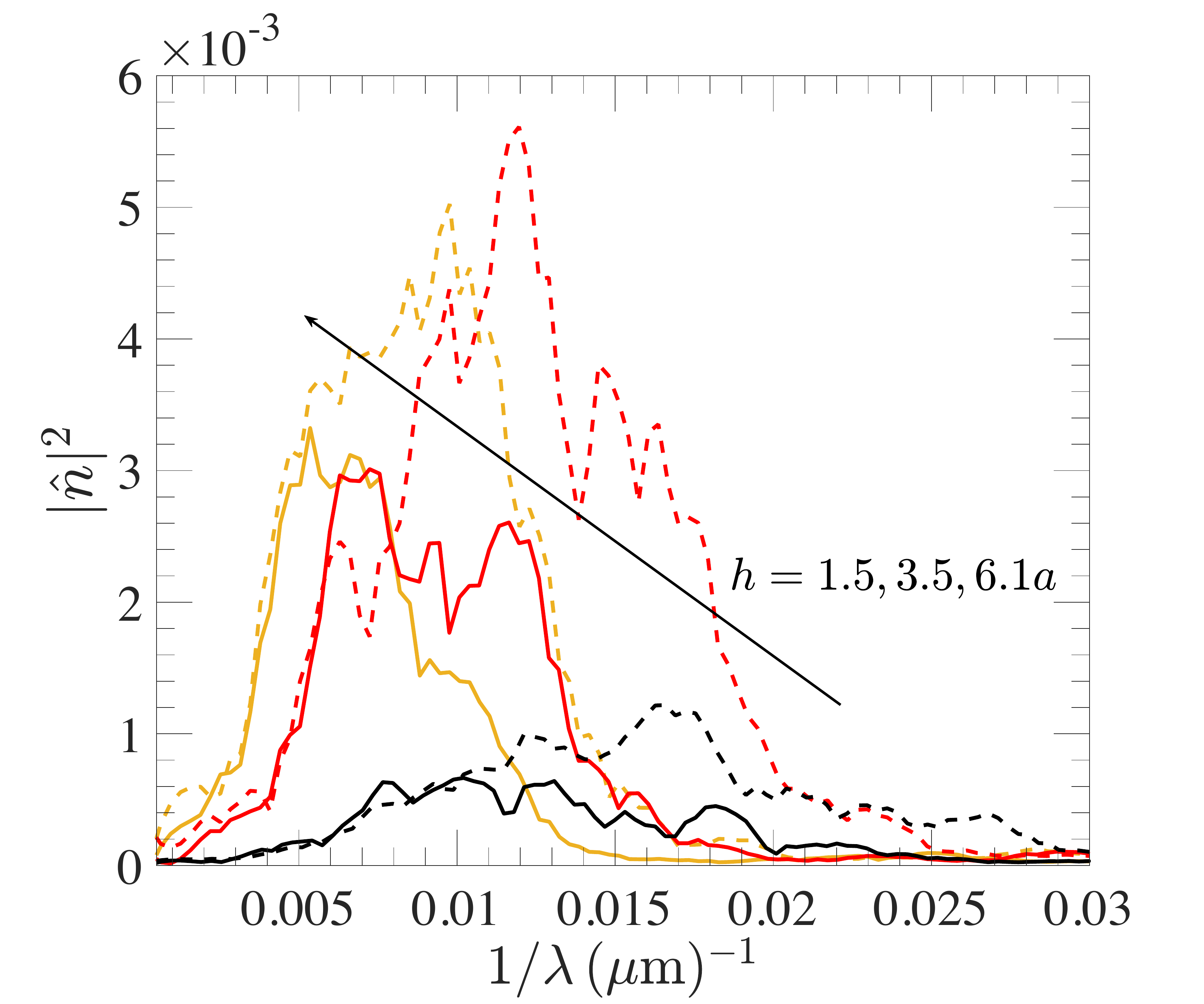}
  \caption{Comparison between deterministic (dashed lines) and
    stochastic (solid lines) simulations at the characteristic time
    $t=t^{\star}$ given in Table \ref{tab:charac_time}. Results are
    shown for three gravitational heights, $h = 1.5a$ (black), $3.5a$
    (red) and $6.1a$ (orange), and averaged over four realizations.
    Left: log-log plot of the height distribution $P(h)$.
    Right: Fourier spectrum $\left| \hat{n}(2\pi/\lambda,t^{\star})
    \right|^2$ of the fluctuations along the front.}
  \label{fig:comparison_height_spectrum}
\end{figure*}

Figure \ref{fig:comparison_histograms} compares the distributions of
particle positions between the stochastic and deterministic
simulations at $t = t^{\star}$ for two gravitational heights, $h_g =
1.5a$ and $h_g = 6.1a$; animated versions of these figures are
available in the Supplementary Material. A visual inspection reveals
that the apparent wavelength of the instability is increased when
thermal fluctuations are included in the system.  \deleted{In Figure
\ref{fig:comparison_height_spectrum} we quantify this effect by
comparing $\left| \hat{n}(k_y,t^{\star}) \right|^2$, and show that the
peak in the Fourier spectrum is shifted toward smaller wavenumbers,
i.e. larger instability wavelengths, when Brownian motion is
included.}  \modified{In Fig. \ref{fig:comparison_height_spectrum}
we quantify this effect by comparing $\left| \hat{n}(k_y,t^{\star})
\right|^2$. At low wavenumbers the amplitude of the spectra is set by
the active flow and is unaffected by the Brownian motion. At high
wavenumbers the Brownian motion damps the amplitude of the spectra
dramatically and this results in an effective shift of the spectra
toward smaller wavenumbers, i.e. larger instability wavelengths, when
thermal fluctuations are included.}  Diffusion also smears out the
front, and the particle distribution $\rho(x)$ is smoother for the
stochastic case at $t=t^{\star}$, see figure
\ref{fig:comparison_histograms}.  
\modified{Likewise, Brownian motion
changes the particle distribution in the direction normal to the wall.
Specifically, the peaks in $P(h)$, created by the active flows, are
shifted toward larger heights and the peak close to the wall is less
pronounced for the stochastic simulations.  The discrepancy with the
exponentially-decaying Gibbs-Boltzmann distribution
\modified{(see $P(h)$ curves at $t=0$ in Fig. \ref{fig:comparison_histograms})}
is due to active motion: the flow field generated in front of a roller close to a wall
lifts the neighboring particles upwards (see Fig. 1c in
\cite{Rollers_NaturePhys}).}  It is also interesting to note that, in
both cases, $P(h)$ is bi-modal when $h_g=1.5a$ and tri-modal when
$h_g=6.1a$. This can be better appreciated in Figure
\ref{fig:comparison_height_spectrum} where the peaks of $P(h)$ are
compared for all three gravitational heights.  The existence of the
third peak for $h_g=6.1a$ is interesting in itself and requires more
investigations to be fully explained. We suspect that enough rollers
are located at the second peak to lift other particles even higher.

These results support our previous observations that the gravitational
height, which sets the initial height of the particles above the
floor, plays a predominant role in the selection of the fastest
growing mode \cite{Rollers_NaturePhys}. Here, as in the experiments,
the gravitational height is controlled by the Brownian diffusion in
the direction perpendicular to the wall.  Note that our initial
conditions are similar, but not identical, to the experiments, where
the particles reach an approximately quasi-steady equilibrium
configuration in a narrow strip.  Our studies also show that Brownian
motion doesn't just control the initial condition but also
quantitatively, though not qualitatively, affects the subsequent
evolution of the fingering instability.  We can estimate a P{\'e}clet
number as $\text{Pe} = v (h_g-a) / D$, where $D=\kt / (6 \pi \eta a)$
is a typical diffusion coefficient and the characteristic speed is
estimated as $v=x_f(t^{\star})/t^{\star}$ where the front position
$x_f(t)$ is the location of the maximum in $\rho(x;t)$. This leads to
values in the range $\text{Pe}\sim12-45$ for our parameters, which
indicates that the motion is dominated by the active flows.  In this
specific example, however, the thermal fluctuations are still
important because they strongly affect the height distribution and the
height determines both the time scale and the wavelength of the
fingering instability.  It is also worthwhile mentioning that our
stochastic simulations are in quantitative agreement with the
experimental measurements. The predominant wavelength in our
stochastic simulations at $t=t^{\star}$ is $\lambda_{\mbox{num}} = 98$
$\mu$m for $h_g = 1.5a$ and $\lambda_{\mbox{num}} = 140$ $\mu$m for
$h_g = 3.5a$, which is in good agreement with the fastest growing mode
measured in the experiments for similar gravitational heights:
$\lambda_{\mbox{exp}} \approx 95$ $\mu$m and $\lambda_{\mbox{exp}}
\approx 160$ $\mu$m respectively \cite{Rollers_NaturePhys}.

\section{Conclusions}
\label{sec:conclusions}

In this paper we presented tools to perform efficient Brownian dynamics
simulations in a half-space.  We proposed a stochastic Adams-Bashforth
method, in which we discretize the deterministic terms using a linear
two-step method, and include the non-trivial stochastic drift using a
random finite difference \cite{BrownianBlobs}.
In Appendix \ref{sec:SPDmobility} we present an
approach for retaining the positive definiteness of the mobility
matrix even if the system reaches an unphysical configuration.
We demonstrated that we can use the
Lanczos algorithm \cite{SquareRootKrylov, RPY_FMM} to generate the
Brownian noise in a small number of iterations independent of the
number of particles. 
Our simulations in Sec. \ref{sec:rollers} illustrate the capacity of
our numerical methods to effectively simulate suspensions of tens to
hundreds of thousands of active Brownian particles confined near a no-slip
boundary over realistic conditions and timescales.  We showed that
Brownian motion affects quantitatively, but not qualitatively, the
development of a fingering instability observed in active roller
suspensions \cite{Rollers_NaturePhys}.  Our PyCUDA codes are publicly
available at
\url{https://github.com/stochasticHydroTools/RigidMultiblobsWall}.

Our simple GPU implementation of the procedure to multiply a vector of
forces by the mobility matrix scales quadratically with the number of
particles. If one wishes to study very large systems with many
hundreds of thousands of particles, one ought to use a linear scaling
method such as the FMM.  It is therefore an important direction for
future work to generalize the point Stokeslet FMM developed in
Ref. \cite{OseenBlake_FMM} to the RPY tensor. Note that in FMM methods
the near-field interactions are computed directly so no special effort
is needed to handle overlapping particles. However, as we discussed in
the body of this paper, handling overlaps with the wall requires
switching from the differential form of the RPY tensor
\eqref{eq:mobilityDifferential}, which is amenable to standard
multipole techniques \cite{RPY_FMM}, to the integral form \eqref{eq:mobilityOverlap}, which
is harder to handle using stnadard multipole expansions.
% Future work should develop efficient methods to 
% approximate \eqref{eq:mobility} (e.g., through pre-tabulation or 
% multipole expansions) either explicitly at the pairwise level, or 
% within an FMM framework at the collective level. 

The method described here only requires a procedure to apply the
mobility matrix to a vector. As such, it can in principle be
generalized to other cases where explicit formulas for the mobility
matrix are available. For example, recently the RP tensor has been
computed for particles confined in a spherical cavity
\cite{SphericalCavity_Zia}.  However, the key difficulty is that the
Lanczos iterative method for computing the Brownian increments is not
going to converge in a constant number of iterations, independent of
the number of particles, unless that hydrodynamic interactions are
strongly screened. Somewhat suprisingly, for particles confined to a
slit channel with no-slip walls \cite{StokesianDynamics_Slit}, the
hydrodynamic interactions decay \emph{slower} than they do for
particles near a single no-slip boundary; it is known that the decay
is $1/r^2$ instead of the $1/r^3$ as in the case studied here
\cite{HI_Confined_Decay}.  It is an interesting question for future
study to explore how well the Lanczos iterative method converges for
suspensions in a thin slit channel, a situation of great practical
interest.  For bulk three-dimensional suspensions it is known that the
conditioning number of the RPY mobility matrix increases with the
number of particles. For periodic boundary conditions, a new linear
scaling method to compute the Brownian increments has recently been
proposed. In this Positively Split Ewald method \cite{SpectralRPY} one
employs an Ewald method to split the hydrodynamic interactions into
near- and far-field interactions. The near-field interactions are then
handled using the Lanczos method, which converges in a small number of
iterations, and the far-field interactions are treated using
fluctuating hydrodynamics.

In this work we only considered the RPY tensor, which captures the
far-field hydrodynamic interactions but does not accurately model
near-field hydrodynamics. There are several techniques that can go
beyond the RPY tensor. The first approach, which is useful for
spherical particles, is to use multipole expansions of higher
order. This has been done systematically via a Galerkin truncation in
recent work by Singh and Adhikari \cite{Galerkin_Wall_Spheres}; when
truncated at the level of stresslets this approach is closely-related
to the widely-used method of Stokesian dynamics
\cite{StokesianDynamics_Wall}. The Galerkin approach ensures an SPD
mobility matrix so Brownian motion can be added straightforwardly,
however, the key difficulty lies in efficient implementation for
systems of many particles. Another approach is to use boundary
integral methods
\cite{BoundaryIntegral_Periodic3D,BoundaryIntegral_Wall}, which
explicitly discretize the boundary and are thus more general. These
methods do not ensure an SPD matrix, and are also still too expensive
to use for suspensions of thousands of particles. Another issue is
that both multipole expansion and boundary integral methods break down
when particles (nearly) overlap each other or the wall. This makes
handling Brownian motion difficult. A third approach is to represent
colloidal particles as rigid multiblobs constructed by rigidly
connecting ``blobs'' that interact hydrodynamically via the RPY tensor
\cite{RigidMultiblobs,RigidMultiblobs_Swan}. This method requires the
same building blocks as we used here, and thermal fluctuations can be
added in a natural way using the Lanczos method applied to the
blob-blob RPY mobility. In future work we will explore such rigid
multilbob methods for Brownian dynamics near a no-slip boundary.

\section{Supplementary material}
See supplementary material for animated versions of figure
\ref{fig:comparison_histograms} for gravitational heights
$h_g=1.5a,\,3.5a,\,\mbox{and } 6.1a$.

\section{Acknowledgment}
We are grateful to Paul Chaikin and Michelle Driscoll for numerous
discussions regarding the fingering instability in active roller
suspensions, and to Edmond Chow for discussions of the Lanczos method.
This work was supported in part by the National Science Foundation
under award DMS-1418706, and by the U.S. Department of Energy Office
of Science, Office of Advanced Scientific Computing Research, Applied
Mathematics program under Award Number DE-SC0008271. B. Delmotte was
supported partially by the Materials Research Science and Engineering
Center (MRSEC) program of the National Science Foundation under Award
Number DMR-1420073.  We thank the NVIDIA Academic Partnership program
for providing GPU hardware for performing the simulations reported here.

\appendix 
\modified{
\section{SPD mobility matrix}
\label{sec:SPDmobility}

In this Appendix we propose a method to regularize the mobility matrix to allow for
particle-wall overlaps.}
Since overlap with the wall is unphysical there is no unique choice of
how to define the pairwise mobility, and existing literature on the RPY tensor
in the presence of a wall does not address this question \cite{StokesianDynamics_Wall,RPY_Shear_Wall}.
There are two physical requirements, however, that any proposal must satisfy.
The first is that a particle whose center touches the wall must stop moving regardless of what forces
are applied to it or to other particles. This ensures that a particle cannot leave the domain and cross
through the wall even in the presence of noise, because the Brownian
displacement of a particle that is about to leave the wall also vanishes if the particle's mobility vanishes.
More precisely, we want the self-mobility
$\bMtt_{ii}$ and the cross-mobilities ${\bMtt}_{ij}$ go to
zero smoothly as the overlap between particle $i$ and the wall increases.
The second requirement is that the mobility matrix must remain
SPD for all particle configurations.

Yeo and Maxey proposed a strategy to define the pairwise mobility for particles overlapping a boundary in the context of the
Force Coupling Method (FCM) \cite{ForceCoupling_Channel}, and this was subsequently
generalized to more general types of boundary conditions in the
context of the Immersed Boundary method
\cite{RigidIBM,RigidMultiblobs}.  The idea is to replace the integrals
over the particle surface in Eq. \eqref{eq:mobilityOverlap} by volume
integrals on the physical domain (half-space above the wall) $\Omega$
and to subtract the effect of an image particle located beyond the
wall, leading to the general definition of the Rotne-Prager-Yamakawa (RPY) tensor in the presence of a wall,
\eqn 
\label{eq:mobility}
\bs{\mathcal{R}} \left( \bq_i, \bq_j \right) =  \fr{1}{(4\pi a^2)^2}
\int_{\Omega} d\bs{x} \int_{\Omega} d\bs{y} \; \bs{G}(\bs{x},\bs{y}) \\
\cdot \left( \delta(|\bs{x} - \bq_i|-a) - \delta(|\bs{x} - \bq^{im}_i|-a)\right)\nonumber \\
\cdot \left( \delta(|\bs{y} - \bq_j|-a) - \delta(|\bs{y} - \bq^{im}_j|-a)\right)\nonumber
\eqnend 
where $\delta(x)$ is the Dirac delta function and the image particle is
located at 
$\bq^{im}_i = \bq_i - 2 \hat{\bs{e}}_z(\hat{\bs{e}}_z \cdot \bq_i)$,  
where we take the $z$ axis to be perpendicular to the wall and directed into the domain.
Unfortunately, the expression \eqref{eq:mobility} involves complicated
integrals that cannot, to our knowledge, be computed analytically.

Instead, as an alternative to \eqref{eq:mobility} in this paper we use an \emph{ad hoc} approximation
to define a mobility that satisfies the physical requirements yet is trivial to compute. We take
\eqn
\label{eq:effMob} 
\bMtt = \bs{B} \wtil{\bMtt} \bs{B},
\eqnend 
where $\bs{B}$ is a diagonal matrix with diagonal blocks
\eqn 
\bs{B}_{ii} = H_0\left( z_i / a \right) \II,
\eqnend 
where the smoothed Heaviside function $H_0$ is
\eqn 
H_0(x) = \left\{ \begin{array}{cl} 
1 & \;\mbox{ if } x > 1,\\ 
x & \;\mbox{ if } 0 \le x \le 1\\
0 & \;\mbox{ if } x < 0.
\end{array} \right.  
\eqnend 
The regularized mobility matrix $\bs{\wtil{M}}$
is the mobility given \eqref{eq:mobilityOverlap} but 
always evaluated for configurations in which neither particle overlaps the wall,
\eqn
\wtil{\bMtt}_{ij}=\bs{\mathcal{R}} \left( (x_i,y_i,\max(z_i,a)), \, (x_j,y_j,\max(z_j,a)) \right).
\eqnend
As desired, the mobility matrix defined by \eqref{eq:effMob} and \eqref{eq:mobility}
is SPD for all configurations and its sub-blocks ${\bMtt}_{ij}$ smoothly go to zero as
either particle $i$ or $j$ overlaps the wall. 
We note that the \emph{ad hoc} approximations like Eq. \eqref{eq:effMob} may
be useful for other Langevin equations where the system can reach an
unphysical state for which the mobility is either undefined or not SPD, or both.
This approach is much simpler and more efficient than trying to strictly prevent
unphysical configurations, as in Metropolis schemes \cite{MetropolizedBD}.
\modified{It is important to ensure, however, that the Heaviside regularization
is only applied infrequently, and therefore does not affect
the results significantly. We ensure this here by including a repulsive potential \eqref{eq:soft_potential} with the wall.}

\modified{
\section{Temporal accuracy}
\label{sec:accuracy}

In this Appendix, we validate and test the accuracy of our numerical schemes on equilibrium systems
by comparing the equilibrium distribution sampled by the BD with the desired Gibbs-Boltzmann distribution,
which we can sample ``exactly'' using a standard
Markov Chain Monte Carlo (MCMC) method \cite{Allen_Tildesley_book}.
}
Specifically, we examine here the distribution $P(h)$  of particle heights $h$ above the wall,
as well as the distribution of pairwise distances given by the pair correlation function $g(r)$.
Because the system is quasi-two-dimensional, we define the radial
distribution $g(r)$ using the three-dimensional distance
$r=(x^2+y^2+z^2)^{1/2}$, but normalize the probability density function as if the suspension were two-dimensional.
This ensures that $g(r)$ goes to unity for $r \gg a$ and that $g(r<2a)=0$ for a hard-sphere suspension.
In the equilibrium simulations we use pseudo-periodic boundary conditions
(PPBC) in the two directions parallel to the wall. We compute the
forces $\bF$ using the minimum image convention between particles as
with standard Periodic Boundary Conditions. 
To account for the longer range of the hydrodynamic interactions we
consider hydrodynamic interactions between particles in the unit cell
and in the first eight neighbor cells, similarly to what is done
in \cite{Osterman2011}.
We can reduce the truncation error in the
hydrodynamic interactions by including the effect of particles in
further neighbor cells (see Ref. \cite{Osterman2011} for some fast approximations);
because the screened hydrodynamic interactions decay fast with
the distance this sum is absolutely convergent. A more sophisticated
approach will be to use a variation of the Ewald's summation method 
\cite{RegularizedStokeslets_WallPeriodic}. However, for
validating the scheme we do not need to refine our approximation
because in the Stokes regime the equilibrium distribution is {\emph not}
affected by the hydrodynamic interactions. The only necessary requirement is
to use the same mobility in the stochastic and deterministic parts of
the schemes to obey the fluctuation dissipation balance.
In the numerical experiments of section \ref{sec:rollers} we did not use PPBC
since the experiments \cite{Rollers_NaturePhys} are performed in a finite sample and not in a bulk suspension.

\begin{figure}
\includegraphics[width=0.49 \textwidth]{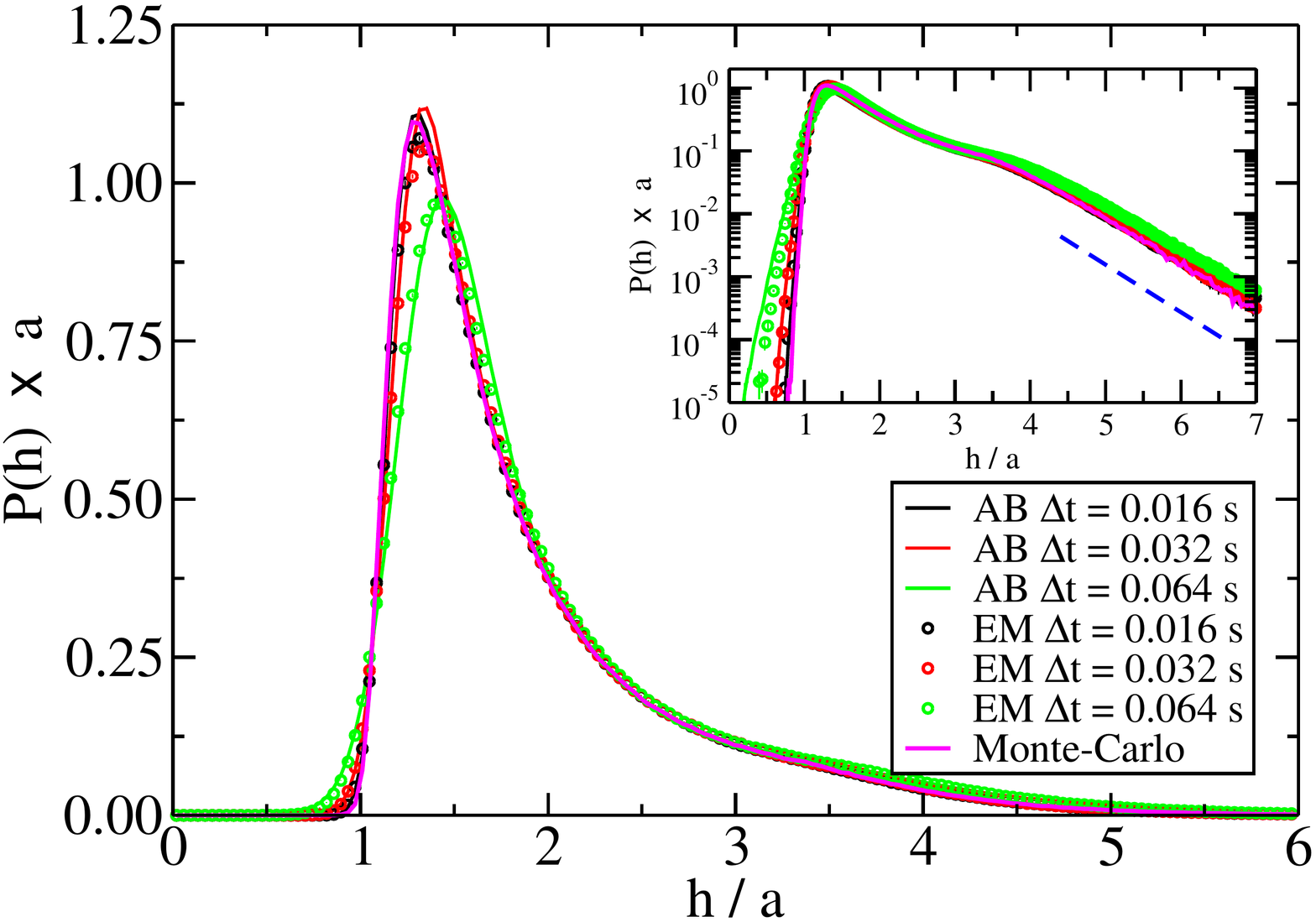}
\includegraphics[width=0.49 \textwidth]{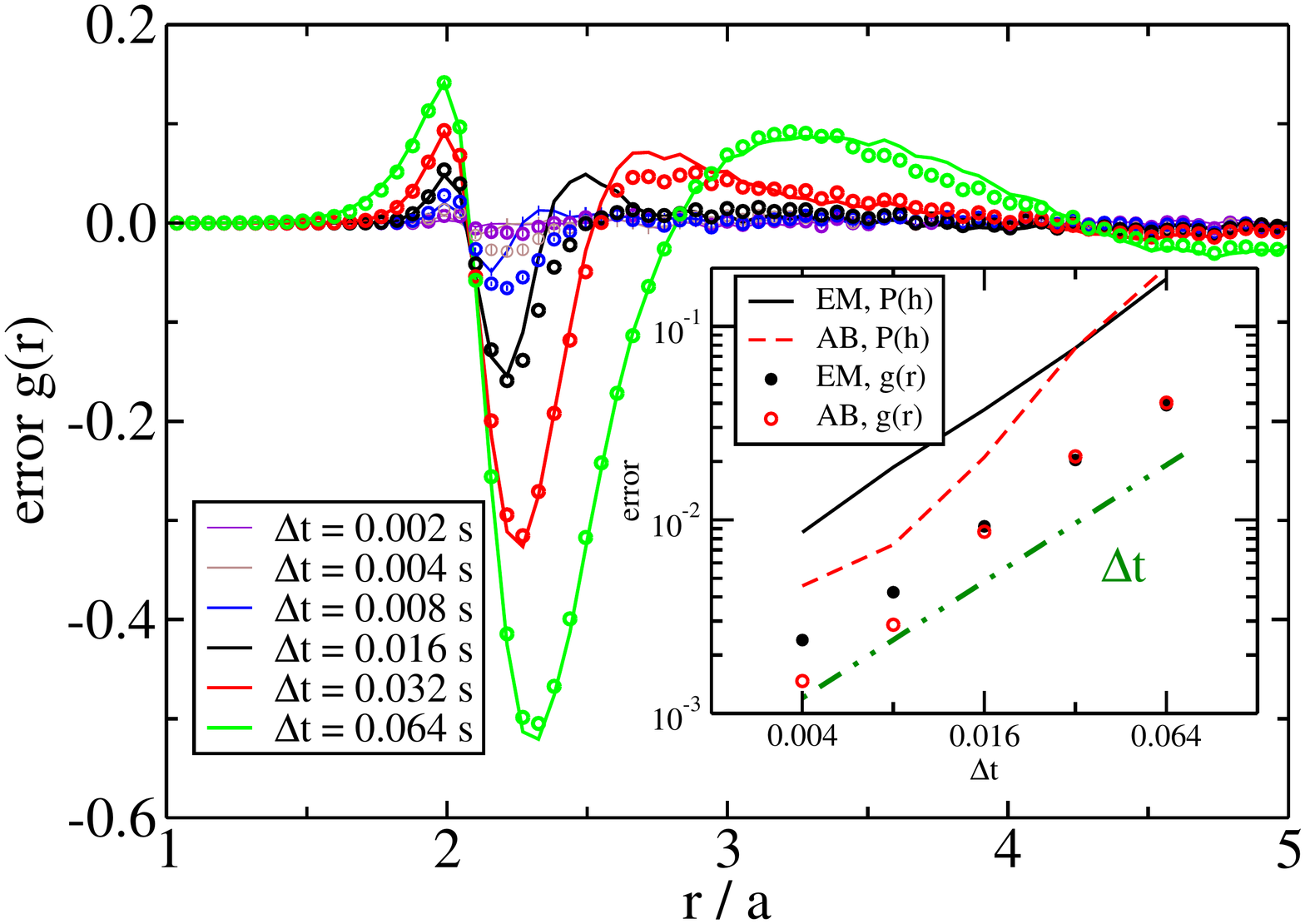}
\caption{\modified{
Stochastic accuracy of the Euler-Maruyama and Adams-Bashforth schemes
for several time step sizes, for a pseudo-periodic passive suspension of $N=1000$ particles,
with area fraction $\phi=0.25$, $U_0=4\kt$, $b=0.1a$ and $h_g=1.58a$.
\modified{The left panel compares the histogram of particle heights $P(h)$
to the ``exact'' distribution sampled by a Monte-Carlo method.
The inset shows the 
exponential decay $\sim\exp(-mgh/\kt)$ (dashed line) due to the gravity field, as well as the}
increased probability of particle-wall overlaps when
$\dt$ is larger than the steric time $\tau_U=0.032\,s$.
The right panel shows the error in the radial distribution function
$g(r)$ with respect to the exact result (computed with a Monte-Carlo method)
for the EM scheme (dots) and AB scheme (lines). \modified{The inset shows
the $L_2$ norm of the error in $P(h)$ and $g(r)$ versus the
time step size, demonstrating the first-order weak accuracy of both methods}.
We can see that for sufficiently small time step sizes AB is more
accurate than the EM scheme.
}}
\label{fig:dt}
\end{figure}

\modified{In Fig. \ref{fig:dt} we show $P(h)$ and the error in $g(r)$
(recall that the scale for $g(r\gg a)\approx 1$)
for BD performed using the EM \eqref{eq:EM} and AB \eqref{eq:AB} schemes, 
for several time step sizes $\dt$ and the parameters given in Section \ref{sec:rollers}.}
The tolerance of the Lanczos algorithm is set to $\epsilon_m = 10^{-4}$.
The fact that the correct equilibrium GB distribution is obtained for small $\dt$
indicates that fluctuation-dissipation is preserved by our temporal integration schemes
because the stochastic drift term is correctly captured by the RFD.
For large time step sizes compared with the steric
characteristic time, such as for example $\dt = 2 \tau_U = 0.064\,s$, the
results show a large deviation with respect to the Monte Carlo results. 
We can see also that the probability of overlap between
particles and the wall becomes non-negligible.  However, when the time step size is
not too large compared with the steric characteristic time, for example
$\dt \le 0.5 \tau_U = 0.016\, s$, the agreement between MCMC
and BD simulations is quite good.
The average/overall error in the histograms as a function of $\dt$
is shown in the inset in the right panel in Fig. \ref{fig:dt}.
\modified{We see that the AB scheme \eqref{eq:AB} is
more accurate than the EM scheme \eqref{eq:EM}
except for the time step sizes that do not resolve the relevant time scales, in which case both schemes are similar.}

\begin{comment}
\begin{figure}
\includegraphics[width=0.49 \textwidth]{gr_pseudo2D_error_run411.pdf}
\caption{Error in the radial
distribution (pair-correlation) function $g(r)$ for the Euler-Maruyama
and Adams-Bashforth schemes and several time step sizes for a pseudo-periodic suspension of $N=1000$ particles,
area fraction $\phi=0.25$, $U_0=4\kt$ and $b=0.1a$.}
\label{fig:dt}
\end{figure}
\end{comment}

%\bibliography{biblio,../References}
% \bibliographystyle{abbrv}
% \bibliographystyle{apsrev}
%\bibliographystyle{ieeetr}
% \bibliographystyle{siam}
%\bibliographystyle{unsrt}

\end{document}